\documentclass[
preprintnumbers,
 amsmath,amssymb,
 aps,
]{revtex4-2}

\usepackage[margin=1.3cm]{geometry}
\usepackage{capt-of}

\usepackage{graphicx}
\usepackage{dcolumn}
\usepackage{bm}

\usepackage{color}

\begin{document}

\title{Tricritical behavior in epidemic dynamics with vaccination}

\author{Marcelo A. Pires$^{1}$}
\thanks{marcelo.pires@delmiro.ufal.br}

\author{Cesar I. N. Sampaio Filho$^{2}$}

\author{Hans J. Herrmann$^{2,3}$}

\author{Jos{\'e}  S. Andrade Jr.$^{2}$}

\affiliation{$^{1}$Universidade Federal de Alagoas, Campus do Sert{\~a}o, 57480-000, Delmiro Gouveia - AL, Brasil \\ 
$^{2}$Universidade Federal do Cear{\'a}, Campus do Pici, 60451-970, Fortaleza - CE, Brasil \\
$^{3}$PMMH, ESPCI, 7 quai St. Bernard, 75005 Paris, France
}

\begin{abstract} 
We scrutinize the phenomenology arising from a minimal vaccination-epidemic (MVE) dynamics  using  three methods: mean-field approach, Monte Carlo simulations, and  finite-size scaling analysis. The mean-field formulation  reveals that the MVE model exhibits either a continuous or a discontinuous active-to-absorbing phase transition, accompanied by bistability and a tricritical point. However, on  square lattices, we detect no signs of bistability, and we disclose that the active-to-absorbing state transition has a scaling invariance and critical exponents compatible with the continuous transition of the directed percolation universality class. Additionally, our findings indicate that the tricritical and crossover behaviors of the MVE dynamics 
 belong to the universality class of mean-field tricritical directed percolation.

\url{https://www.sciencedirect.com/science/article/abs/pii/S0960077923006628}
 
\end{abstract}

\maketitle


\section{Introduction}\label{sec:intro}

The current pandemic revealed that disease spreading still poses several challenges for society. In order to combat an epidemic spreading, vaccination is  a great ally~\cite{wang2016statistical}. As many social and biological processes, epidemic and vaccination dynamics  can be addressed with tools from statistical physics due to the stochastic nature of the disease spreading~\cite{pastor2015epidemic,serafino2022digital,ponte2021tracing,s2022spatio}.

From the statistical physics point of view, 
 epidemic processes undergo nonequilibrium phase transitions~\cite{marro2005nonequilibrium,hinrichsen2000non}.  
Nonequilibrium physics is not a simple extension of equilibrium physics, but is associated with new phenomena.
There exist many epidemic models~\cite{wang2016statistical,pastor2015epidemic} and the  SIS~(Susceptible-Infected-Susceptible) is the paradigmatic model that presents a phase transition from an active phase, where there is a self-sustaining perpetuation of the disease in the population, to an absorbing phase, where the disease is eradicated. 
The nonequilibrium phase transition of the SIS model is continuous (second-order) and follows the behavior of the  important directed percolation (DP)  universality class of absorbing-state transitions~\cite{hinrichsen2000non}. The DP universality class includes not only epidemic processes, but a  large variety of spreading phenomena observed in Nature~\cite{marro2005nonequilibrium}.

As aforementioned, vaccination is a key tool to mitigate the impact of a disease spreading. In mathematical modeling, vaccination can be minimally coupled to the SIS models by introducing the vaccinated compartment, $V$, with the corresponding rates. In such case, vaccination induces an increase in the epidemic threshold, reducing the impact of a self-sustaining chain of contagion.
This effect preserves the continuous nature of the SIS phase transition as shown, for instance, in~\cite{zhou2003stability,shaw2010enhanced,pires2017dynamics,castro2022effect}.
Nevertheless, due to the fact that all vaccines have a limited effectiveness, the nature of this transition can change if a vaccinated individual becomes infected, passing directly from compartment $V$ to $I$~\cite{kribs2000simple}.
Following this work, subsequent models for epidemics with vaccination  exhibiting  both continuous and  discontinuous transitions have been extensively studied in the mathematical community by means of the
forward-backward bifurcation theory~\cite{kribs2002vaccination,arino2003global,brauer2004backward,sharomi2007role,reluga2007resistance,buonomo2011backward,gerberry2016practical,zhang2018vaccination,nudee2019effect,lacitignola2021managing,rashkov2021complexity,saldana2021modeling,song2021basic}. 
\textcolor{black}{However, the overwhelming majority of such studies have focused on deterministic models considering fully-mixed populations. Here, we investigate the statistical and spatial properties of a minimal vaccination-epidemic model by means of three methods: 
Monte Carlo simulations, finite-size scaling (FSS)  and mean-field approach.
 As we will see,  our findings revealed the presence of continuous and discontinuous phase transitions depending on the system's dimensionality.}

This work is organized as follows. In section II, we describe  our stochastic minimal vaccination-epidemic (MVE) model with the corresponding mean-field treatment.  In section III, we present the results and discussion from Monte Carlo (MC) simulations and scaling theory. Final remarks are presented in section IV.

\section{Model}\label{sec:model}

\begin{figure}[!htb]
 \centering
\includegraphics[width=0.35\textwidth]{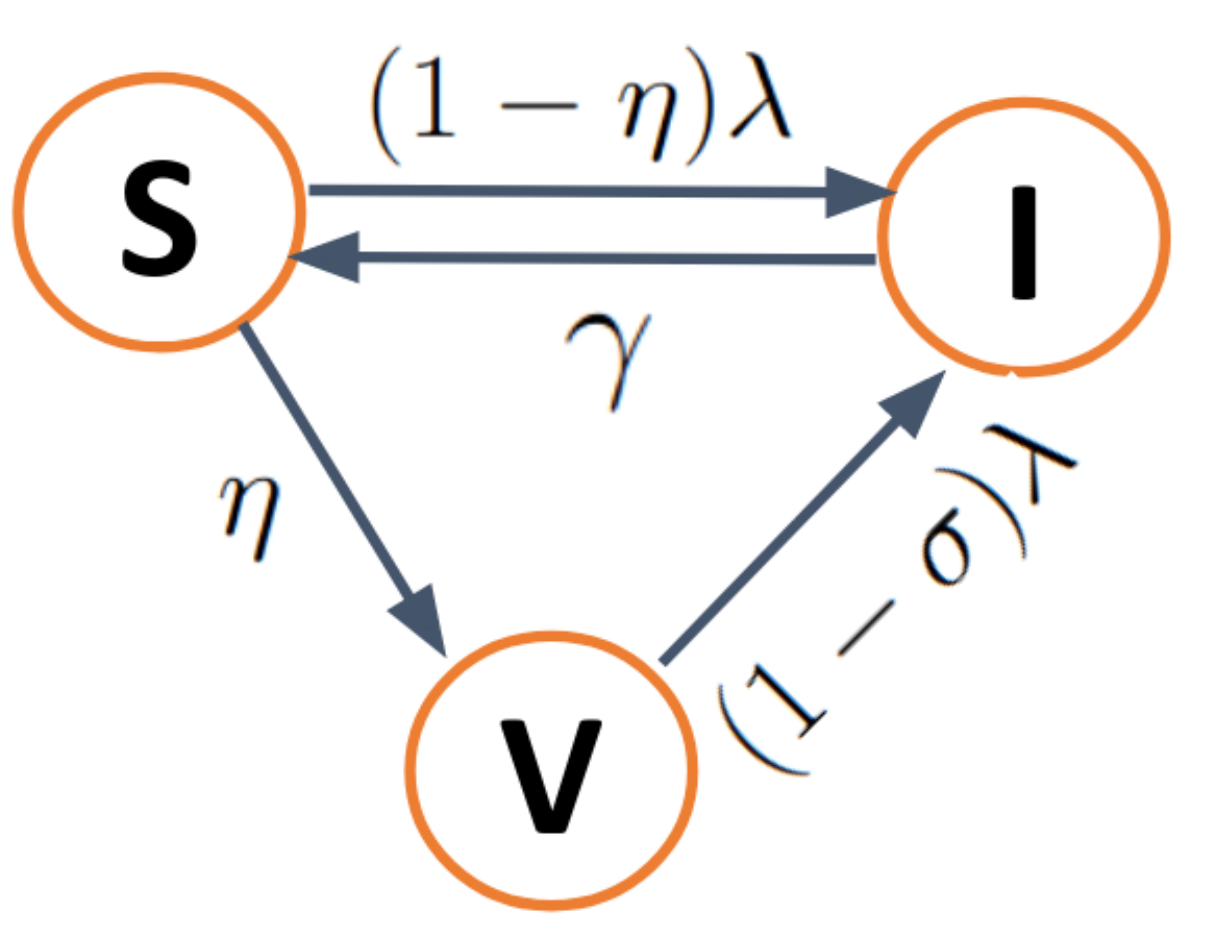}
\caption{Sketch of the vaccination dynamics with an illustration of the probabilities $\{\lambda,\eta,\gamma,\sigma\}$ described in the text.  For brevity, we call this model as minimal vaccination-epidemic (MVE) dynamics.  \textcolor{black}{The MVE model is composed by four compartmental transitions: (i) two activated by a pairwise interaction ($S+I \stackrel{}{\rightarrow} 2I$ and $V+I \stackrel{}{\rightarrow} 2I$);  and (ii) two non-mediated by interactions ($S \stackrel{}{\rightarrow} V$ and $I \stackrel{}{\rightarrow} S$). }  }
\label{fig:min_vac_sudden_model_sketch}
\end{figure} 

\textcolor{black}{Minimal models can be highly instructive in providing insights that can help to elucidate the mechanisms inherent in the large-scale behavior of a given system~\cite{siegenfeld2020models}. }
In this sense, we consider a statistical and minimal version of the model presented in  Ref.~\cite{reluga2007resistance}. 
Our model is sketched in Fig.~\ref{fig:min_vac_sudden_model_sketch}.
  The $N$ individuals of a population can be divided into $3$ states: Susceptible ($S$), Infected ($I$) and vaccinated ($V$).  If a given individual $i$ is susceptible, 
a vaccination event $S \stackrel{}{\rightarrow} V$ takes place with probability $\eta$. If $i$ does not get vaccinated (event that takes place with probability $1-\eta$) then  $i$ becomes infected with probability $\lambda_{ef}$.
For a full graph $\lambda_{ef} = \lambda I/N$,  whereas for structured graphs we need to consider  the number of infected neighbors, $n_I(i)$. For the square lattices $\lambda_{ef} = \lambda n_I(i)/4$ 
and for the random $k$-regular graphs $\lambda_{ef} = \lambda n_I(i)/k$. 
If an individual $i$ is already vaccinated and meets an infected peer, then a contagion event takes place with probability $(1-\sigma)\lambda_{ef}$, where $\sigma$ is the vaccine effectiveness. If $\sigma=1$ the vaccine has $100$\% of effectiveness. 
In turn, the infected individuals recover becoming susceptible again, $I \stackrel{}{\rightarrow} S$,  with probability $\gamma$. We assume that the virus mutates fast having many variants so that an infection is not sufficient to effectively induce immunity. For this reason, we ignore any transition from compartment $I$ to $V$.

\subsection{Mean-field approach}

From the previously stated concepts and rules, as 
illustrated in Fig.~\ref{fig:min_vac_sudden_model_sketch}, we obtain the mean-field equations:
\begin{equation}\label{Eq:S1}
\frac{dS }{dt} =  
- \eta S  -(1-\eta)\lambda S\frac{I}{N}  +   \gamma I
\end{equation}
\begin{equation}\label{Eq:I1}
\frac{dI}{dt} =  
          (1-\eta)\lambda S\frac{I}{N}  +
(1-\sigma)\lambda V\frac{I}{N} 
          -  \gamma I 
\end{equation}
\begin{equation}\label{Eq:V1}
\frac{dV}{dt} = 
\eta S -  (1-\sigma)\lambda V\frac{I}{N}, 
\end{equation}
\noindent where we denote by $S$, $I$ and $V$ the number of individuals in the compartments $\{S,I,V\}$, respectively. 
Adding the Eqs.~(\ref{Eq:S1}-\ref{Eq:V1}) we obtain $\frac{dS}{dt}+\frac{dI}{dt}+\frac{dV}{dt}=0$ as expected since $N=S+I+V$ is assumed constant.

The Eqs.~(\ref{Eq:S1}-\ref{Eq:V1}) accompanied by $N=S+I+V$ provide a steady-state governed by a third-degree polynomial 
\begin{equation} \label{Eq:I_polinomio}
\rho_{\infty} \left( A\rho_{\infty}^2 + B \rho_{\infty} - C \right) = 0,    
\end{equation}
where $\rho_{\infty}=I_{\infty}/N$ and 
\begin{align} 
 A &=  (1-\sigma) \lambda (1-\eta)     \label{Eq:sol3_A}
\\ B &= (1-\sigma)\left( \gamma  + \eta -\lambda (1 - \eta) \right) 
\\ C &= \eta \left(  1-\sigma - \frac{\gamma}{\lambda}  \right).  \label{Eq:sol3_C}
\end{align}

The $3$ solutions of Eq.~\ref{Eq:I_polinomio} are:
\begin{align} \label{Eq:sol3_0}
\rho_{\infty}^o &= 0
\\ \rho_{\infty}^{\pm} &= \frac{-B}{2A} \pm \sqrt{ \frac{C}{A} + \left(\frac{B}{2A}\right)^2 }. \label{Eq:sol3_Iplus}
\end{align}

The above equations are very instructive.

First, the Eq.~\ref{Eq:sol3_A} shows that  $A\geq 0$ for any choice of the epidemic parameters that are bounded between 0 and 1. This observation simplifies the stability analysis. The  solution  $\rho_{\infty}^o = 0$ corresponds to an absorbing state, the disease-free phase. From a linear stability analysis one finds that $\rho_{\infty}^o$ is stable for  $C>0$ and unstable for $C<0$. At $C=0$ we obtain the critical point $\lambda_c=\frac{\gamma}{1-\sigma}$. The system undergoes a continuous active-absorbing-state phase transition at $\lambda = \lambda_c$, if $B>0$ where the solution $\rho_{\infty}^{+}$ becomes stable. 
In this case, close to the critical point (critical region) we find that the order parameter behaves as $ \rho_{\infty} \sim \left(\lambda - \lambda_c \right)^{\beta_{DP}} $, with $\beta_{DP}=1$. 
If $B<0$ the transition becomes discontinuous and there is a region $\lambda_c \leq \lambda \leq \lambda^{*}$ characterized by a bistability between the solutions   $ \rho_{\infty}^o $ and $ \rho_{\infty}^{+}$. 
At $B=0$ we find the tricritical point (TCP)  $\lambda_T= \frac{\gamma+\eta}{1 - \eta}$. 
At the TCP  $\lambda_T = \lambda_c$ thus $\sigma_T =1 - \frac{\gamma(1-\eta)}{\gamma+\eta}$. 
As a consequence, close to the TCP (tricritical region) we find that the order parameter behaves as $ \rho_{\infty} \sim \left(\lambda - \lambda_T  \right)^{\beta_{T}} $ with $\beta_{T}=1/2$. 
Finally, the solution $\rho_{\infty}^{-}$  yields unphysical and unstable results. 
For further reference,  we summarize the location of the TCP,
\begin{equation}\label{eq:tcp}
(\lambda_T,\sigma_T) = \left( \frac{\gamma+\eta}{1-\eta},  \frac{ \gamma+\eta + \gamma(1-\eta)}{\gamma+\eta} \right).    
\end{equation}

Second, if $A=0$ the higher-order term in  Eq.~\ref{Eq:I_polinomio} disappears which in turn removes any possibility for a  discontinuity in the epidemic transition for this model. From  Eq.~\ref{Eq:sol3_A} we see that $A=0$ if the vaccination program is operated with a $100\%$ effective vaccine, 
\emph{i.e.} $\sigma=1$. 
In any case, even if $\sigma<1$ the  abrupt transition to the spreading phase can be suppressed if the vaccination probability is sufficiently high,  $\eta\geq\eta^{\dagger}$. By setting $\lambda_T= \frac{\gamma+\eta^{\dagger}}{1 - \eta^{\dagger}} = 1$ we obtain explicitly the threshold
\begin{equation}\label{eq:eta_threshold}
\eta^{\dagger}=  \frac{1-\gamma}{2} 
\end{equation}

\subsection{Simulation on complete graphs}

We employ the Gillespie's algorithm~\cite{gillespie1977exact} to obtain the time evolution of the model on complete graphs. 
In this algorithm the time to the next event is exponentially distributed and the probability of each event is proportional to its rate. 
We define  $N_S$, $N_I$ and $N_V$ as the number of Susceptible, Infected and Vaccinated individuals, respectively. 
At each time $t$ we compute the rate of each event described in the previous subsection:
 $r_1 = r_{S\rightarrow V} = \eta N_S$, 
 $r_2 = r_{S\rightarrow I} = (1-\eta) \lambda N_S N_I/N$, 
$r_3 = r_{V\rightarrow I} = (1-\sigma) \lambda N_V N_I/N$, 
$r_4 =r_{I\rightarrow S} = \gamma N_I$. 
Next, we compute the sum of all process rates $r_{tot} = \sum_{i=1}^{i=4} r_i $.
We determine the time increment $\delta t$ by generating a random number from an exponential distribution with parameter $r_{tot}$,  \textit{i}.\textit{e}., 
$\delta t \sim \exp{(r_{tot})} $.  The next event $i$ is chosen with  probability directly  proportional to its rate, 
 \textit{i}.\textit{e}.,  
we sample the next event $i$  uniformly with probability  $p_{i} = r_i/r_{tot}$. Such probability distribution ensures that 
$\sum_{i=1}^{i=4} p_i = 1$. Subsequently, the population variables are updated according to the type of event to be selected:
\begin{itemize}
\item [(E1)] If the event $S\rightarrow V$ is chosen, $N_S$ is decreased by one and $N_V$ is increased by one;
\item [(E2)] If the event $S\rightarrow I$ is selected, one unit is subtracted from $N_S$ and one unit is added to $N_I$;
\item [(E3)] If the event $V\rightarrow I$ is sampled, $N_V$ is decreased by one and $N_I$ is increased by one;
\item [(E4)] If the event $I\rightarrow S$ is picked, one unit is subtracted from $N_I$ and one unit is added to $N_S$.
\end{itemize}
These steps are performed while $t\leq t_{max}$.

\subsection{Simulation on structured graphs}

During the simulation of the MVE model 
on 
square lattices and random $k$-regular graphs
we use two lists: 
(a) $\mathcal{L}_I $: list of infected individuals with size $N_I$;
(b)  $\mathcal{L}_S $: list of susceptible individuals with size $N_S$.
Denote $\delta t$ as a time increment associated with a given step in the simulation.
At each step one of the following events can take place:
\begin{itemize}
\item [(E1)] Vaccination of a susceptible individual with probability $ \eta N_S \delta t$. In this case, 
we choose randomly an individual $i$ from $\mathcal{L}_S $.  The state of $i$ changes as  $S \stackrel{}{\rightarrow} V$.
\item [(E2)] Infection of a susceptible individual with probability $  (1-\eta)\lambda  N_I \delta t$. For this event,  we pick at random an individual $i$ from $\mathcal{L}_I$. Then, we choose randomly a neighbor $j$ of $i$. If $j$ is susceptible, the state of such  individual changes as  $S \stackrel{}{\rightarrow} I$.
\item [(E3)] Infection of a vaccinated individual with probability $  (1-\sigma)\lambda   N_I \delta t$. In this event, we sample uniformly an individual $i$ from $\mathcal{L}_I$. Then, we select at random a neighbor $j$ of $i$. If $j$ is vaccinated, the state of such  individual changes as  $V \stackrel{}{\rightarrow} I$.
\item [(E4)] Recovery of an infected individual with probability $ \gamma N_I \delta t$. In this case, 
we sample uniformly an individual $i$ from $\mathcal{L}_I $.  The state of $i$ changes as  $I \stackrel{}{\rightarrow} S$.
\end{itemize}
Then, we increase the time $t$ by an amount $\delta t$. From the normalization of the total  probability we obtain 
$ \frac{1}{\delta t} = \eta N_S + (2-\eta-\sigma) \lambda N_I + \gamma N_I $. These steps are performed while $t\leq t_{max}$.

\subsection{Further simulation details}

In epidemic models, the absorbing state refers to the configuration without infected individuals. If this state is reached, the dynamics stops. However, in finite systems the absorbing configuration can always be achieved, even in the supercritical phase. To avoid this issue, we apply a quasi-stationary method to characterize the dynamics of the model. Specifically, we apply a reactivation method~\cite{macedo2018reactivating,sampaio2018symbiotic} 
in which we perform a reinfection of one individual (randomly chosen) every time the dynamics reaches the absorbing configuration.

In our simulations we consider a complete graph (also called full graph or fully-connected network) that is a structure where each vertex (individual) interacts with all others.
We also consider a random $k$-regular graph that is a quenched structure where each node has $k$ neighbors. Finally, we consider a square lattice
of lateral size $L$ with a von Neumann neighborhood  ($4$ neighbors) and periodic boundary conditions.

For the finite-size scaling (FSS)
analysis of the MVE model 
we start all simulations with the system fully occupied with infected individuals and 
we follow the procedures delineated above (see also Section 3.4.3 of Ref.~\cite{hinrichsen2000non}).

By means of stochastic simulations, we compute the fraction of infected agents which is the order parameter $\rho$ of the model
\begin{equation} \label{rho_mc}
\rho = \left\langle \frac{1}{N} \sum_{i=1}^{N} x_{i} \right\rangle, 
\end{equation}
\noindent
where $x_{i}=1$ if $i$ is infected, otherwise $x_{i}=0$ and  $\langle\, ...\, \rangle$ denotes the average over  $n_{samples}=10^2$ samples for each parameter setting.

\begin{figure}[!htb]
    \centering
\includegraphics[width=0.45\textwidth]{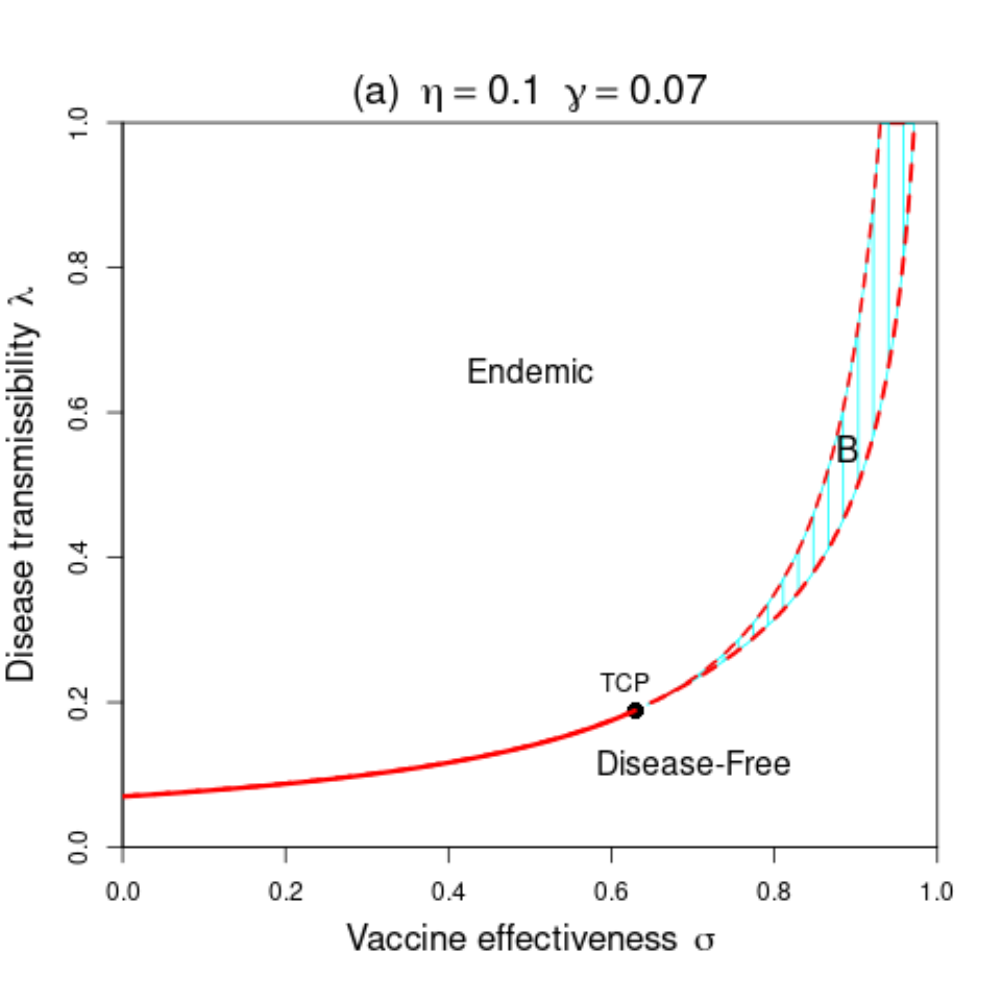}
\includegraphics[width=0.45\textwidth]{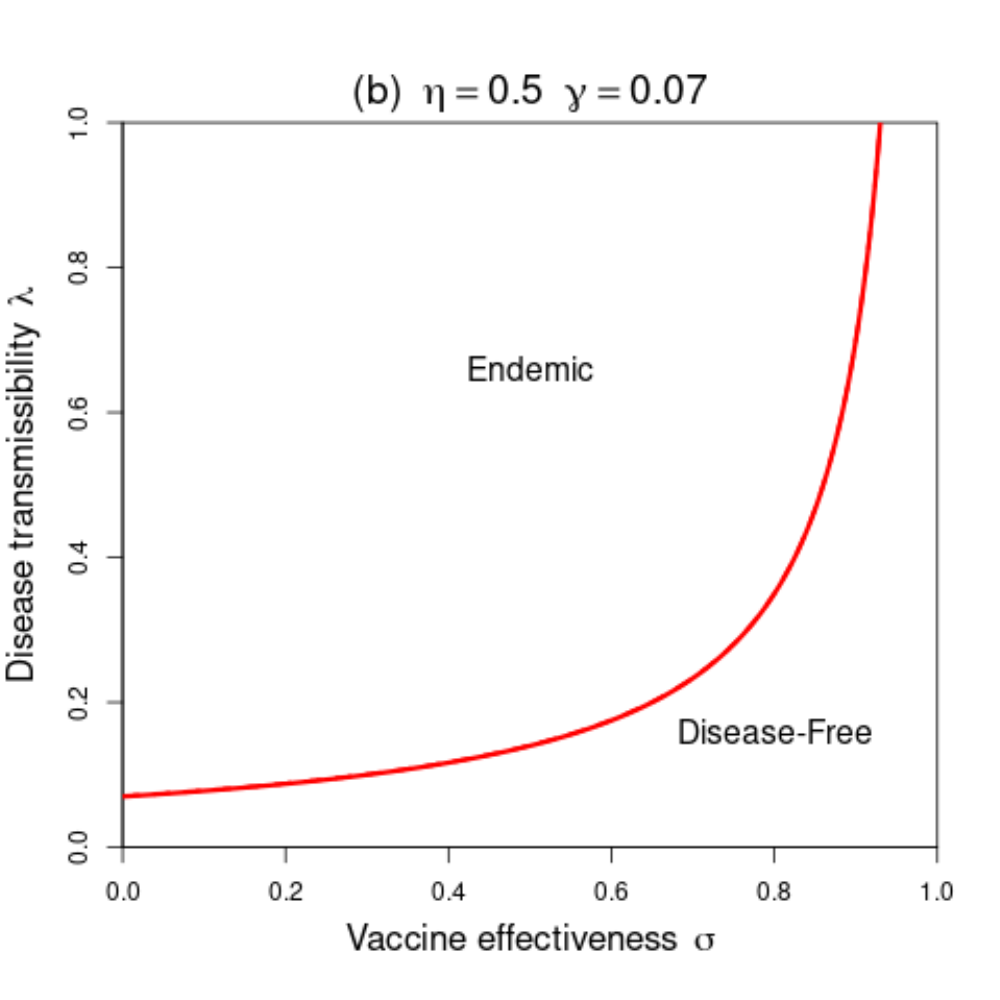}
\caption{Mean-field phase diagram of the minimal vaccination-epidemic (MVE) model, as depicted in  Fig.~\ref{fig:min_vac_sudden_model_sketch}, for $\gamma=0.07$, and: $\eta=0.1$ (a) and 
 $\eta=0.5$ (b). The shaded area corresponds to the bistable (B) phase \textcolor{black}{ where the  dynamics has three fixed points: two stable  ($\rho_{\infty}^{+}$ and $\rho_{\infty}^o$) and one unstable ($\rho_{\infty}^{-}$). In the Disease-Free phase $\rho_{\infty}^o$ is the only the stable fixed point. In turn, $\rho_{\infty}^{+}$ is the unique stable fixed point
in the Endemic phase}. The coordinates of the  tricritical point (TCP) are obtained from Eq.~\ref{eq:tcp}. From the thresholds of  $\rho_{\infty}^{\pm}$ in  Eq.~\ref{Eq:sol3_Iplus}, we obtain the solid and dashed  curves that represent continuous and discontinuous transitions, respectively.  \textcolor{black}{The bistability can be suppressed if the vaccination probability $\eta$ is above the  threshold given by the Eq.~\ref{eq:eta_threshold}.  } }  
\label{fig:steady_state_f1_diagram}
\end{figure}

\section{Results and discussion} \label{sec:res1}

\begin{figure}[!htb]
\centering
\includegraphics[width=0.46\textwidth]{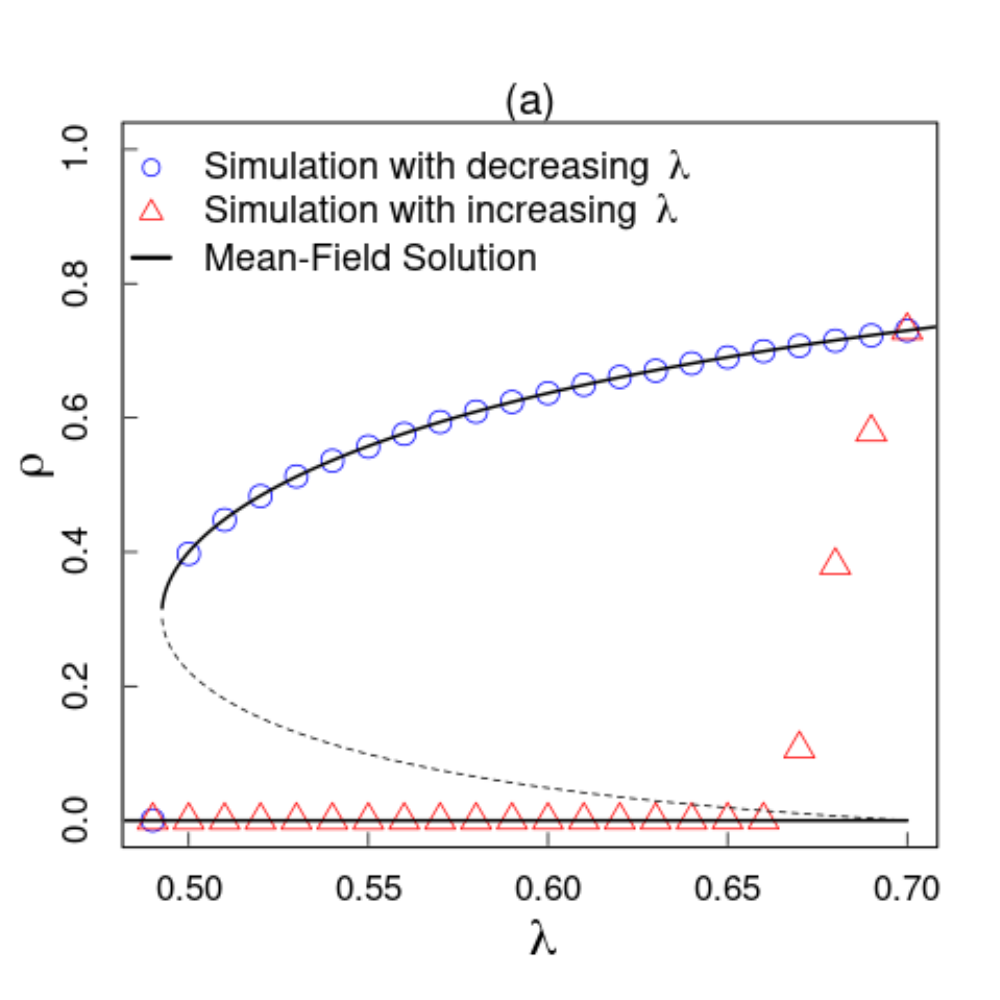}
\includegraphics[width=0.46\textwidth]{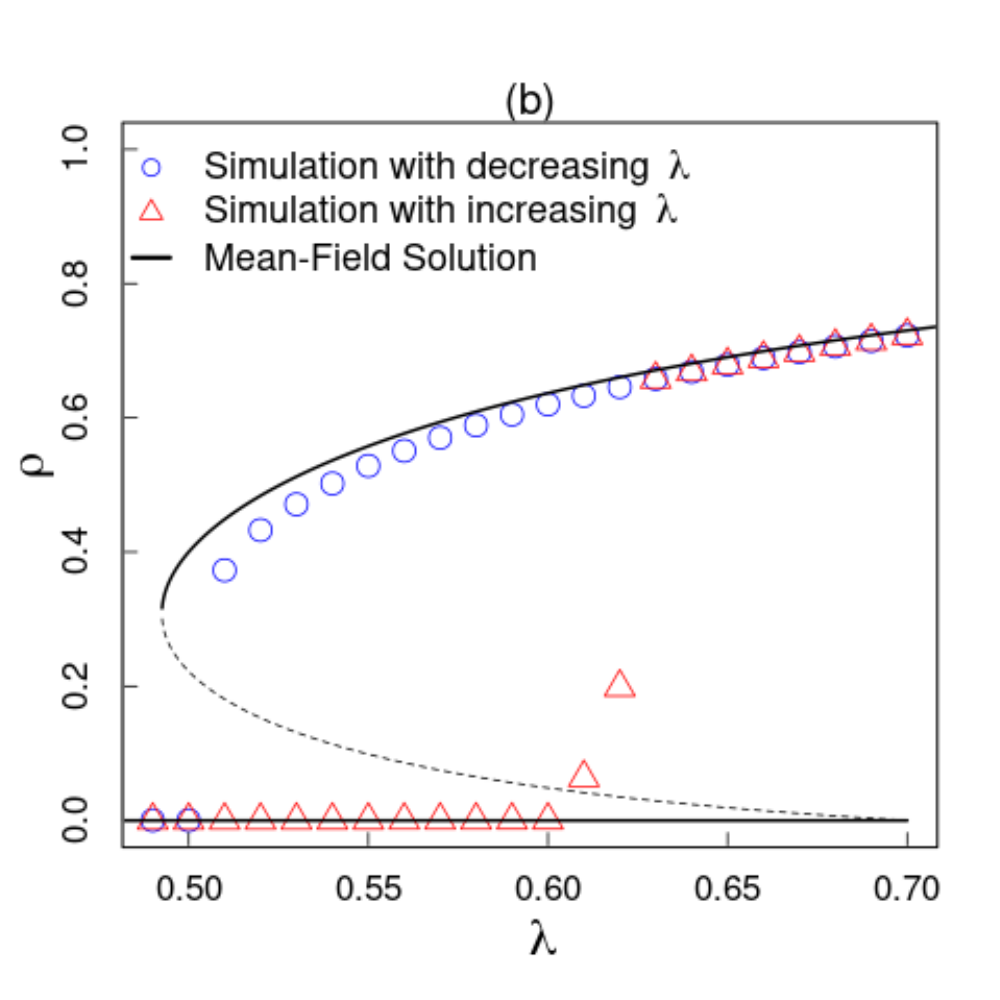}
\includegraphics[width=0.46\textwidth]{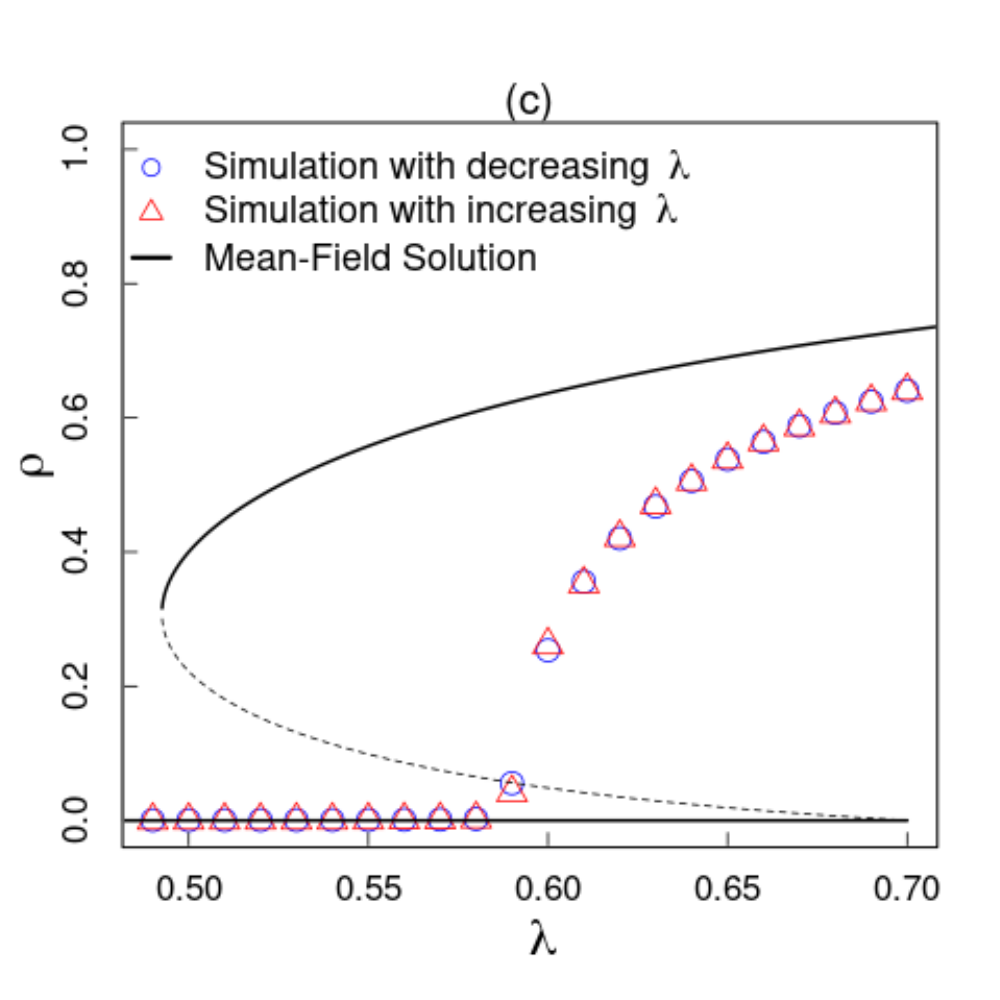}

\caption{Stationary infected density $\rho$ versus the transmissibility $\lambda$ for the MVE model on: (a) complete graphs, (b) random $k$-regular networks with $k=20$, (c) square lattice. Simulations performed with $t_{max}=10^5$ on structures with $N=10^4$ individuals each. The control parameter $\lambda$ is increased (red triangles) and decreased 
(blue circles) in the range $0.49\leq \lambda\leq 0.7$ at constant intervals of $ \Delta \lambda = 0.01$.   \textcolor{black}{In the upper branch the simulations started with $\lambda=0.7$ from $\rho_o=1$ (fully active initial condition), whereas in the lower branch the simulations started with $\lambda=0.49$  from $\rho_o=1/N$ (localized initial condition).  The stable and unstable theoretical solutions, obtained from the Eq.~\ref{Eq:sol3_Iplus}, are represented in the solid and dashed lines, respectively.  } }
\label{fig:hyst}
\end{figure}

\begin{figure}[!htb]
\centering

\includegraphics[width=0.45\textwidth]{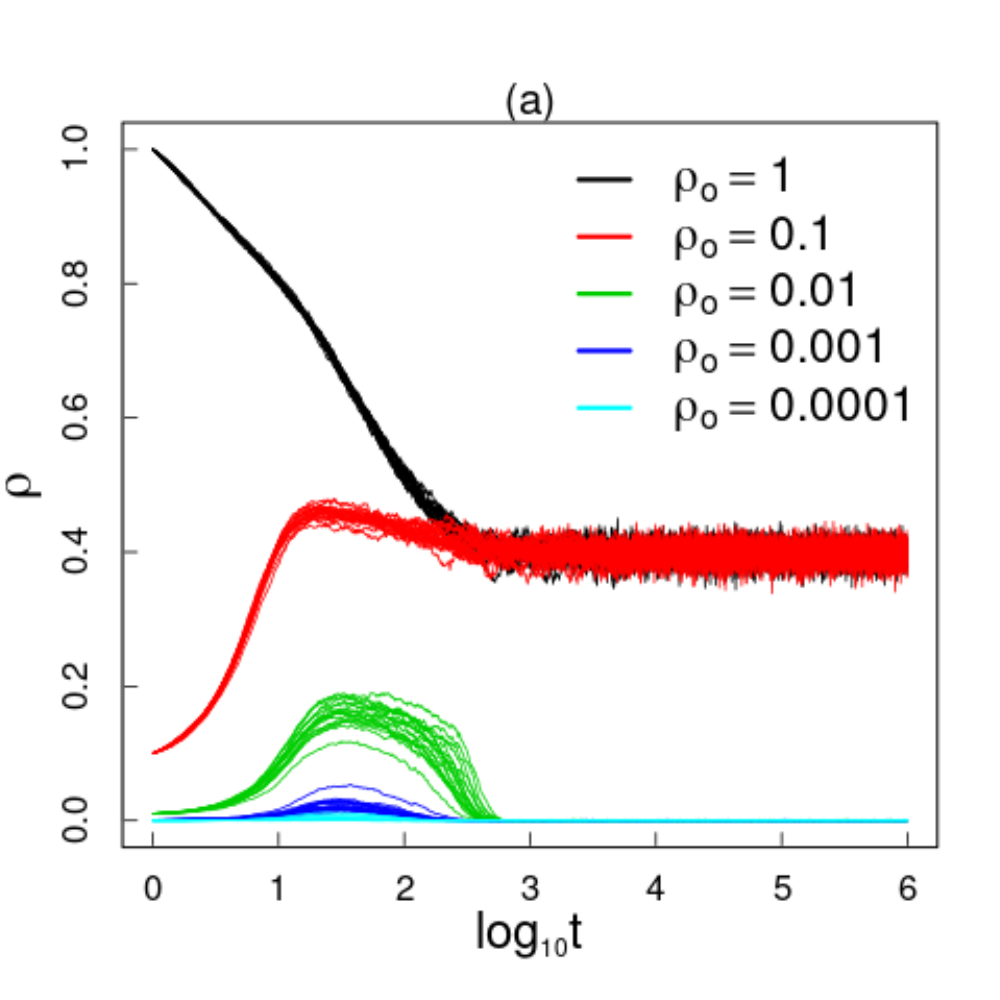}
\includegraphics[width=0.45\textwidth]{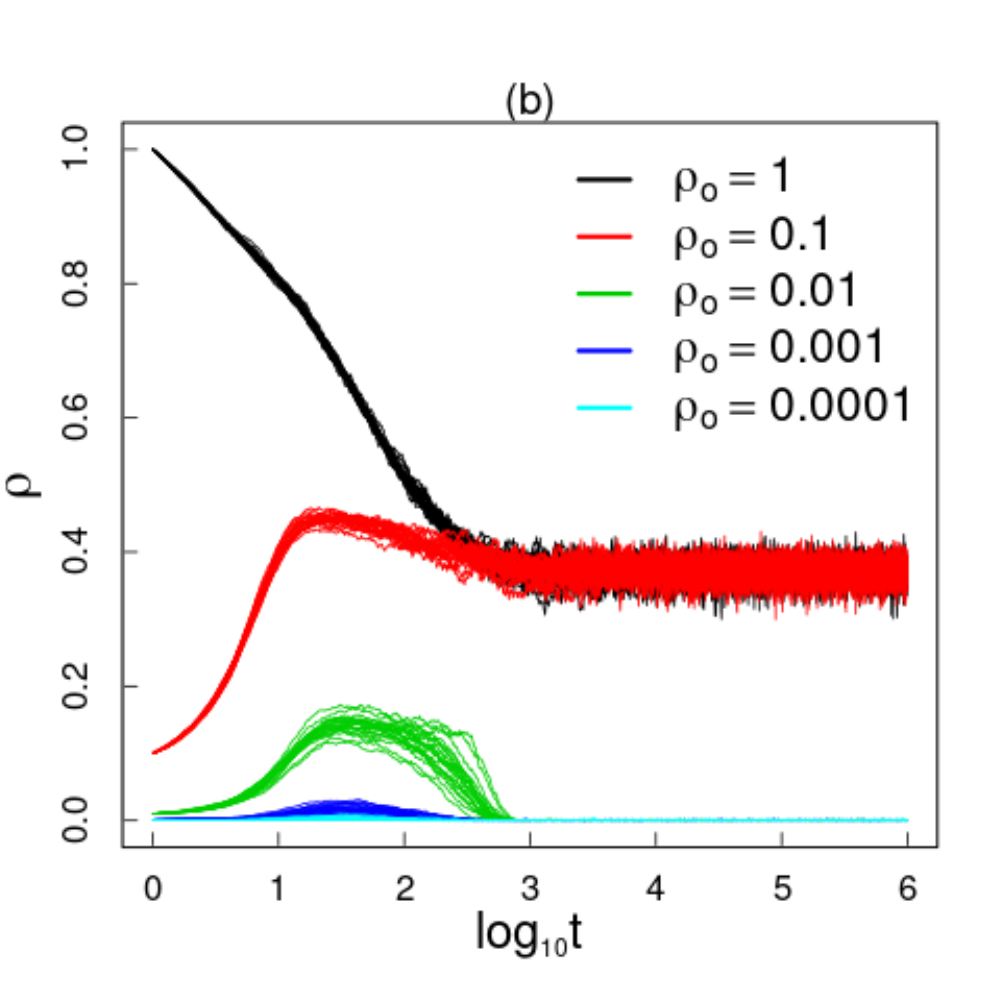}

\includegraphics[width=0.45\textwidth]{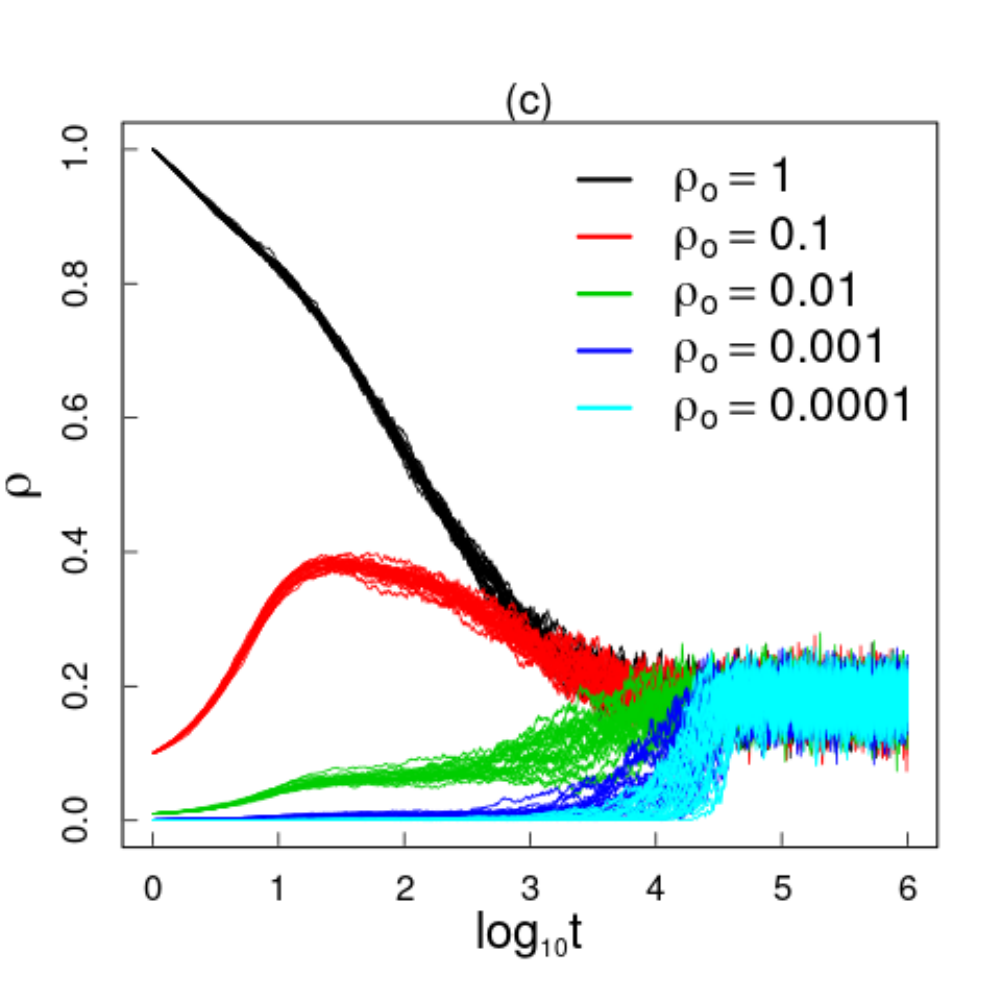}

\caption{Time evolution of the order parameter $\rho$ for several initial conditions $\rho_o$. 
(a) $\lambda=0.50$, complete graphs, 
(b) $\lambda=0.51$, random $k$-regular networks with $k=20$, 
(c) $\lambda=0.595$, square lattice. 
All structures with $N=10^4$ individuals. 
Each panel has $100$ samples, and for better detecting eventual bistability we plot the time series for each parameter setting without averages.
The time series in panels (a-b) exhibit bistable solutions, whereas the panel (c) exhibits a single stable steady state. \textcolor{black}{In panels (a-b) note that in some dynamic scenarios, the presence of pronounced epidemic peaks in the short-term does not necessarily promote persistence of the transmission chain in the long-term. }  }
\label{fig:biestability}
\end{figure}

In this section, we first provide an analysis of the parameter space for the mean-field setup. Next, we focus on the parameter configurations that produce a large discontinuity in mean-field results. That is, 
we set $\gamma=0.07$, $\eta=0.1$ and $\sigma=0.9$
in all the finite-size simulations,  when not stated otherwise.

In Fig.~\ref{fig:steady_state_f1_diagram}, we show two typical  phase diagrams of the MVE model illustrated in 
Fig.~\ref{fig:min_vac_sudden_model_sketch}. In the panel (a)    with $\eta=0.1$, we observe in the phase diagram that the transition from the disease-free phase to the spreading one is continuous if $\sigma\leq\sigma_T$ or discontinuous if $\sigma>\sigma_T$. \textcolor{black}{Such  discontinuity in the order parameter, infected density, is not found in the phase diagram of panel (b) because in this case the vaccination probability  $\eta=0.5$ is greater than the threshold value given by Eq.~\ref{eq:eta_threshold},  $\eta^{\dagger} = (1-\gamma)/2 = 0.465$. That is, prevention of an abrupt spread of an epidemic can be achieved if the vaccination probability is high enough.}

In Fig.~\ref{fig:hyst}, we show the 
long-time behavior of the order parameter $\rho$ versus the control parameter $\lambda$ considering three structures: complete graphs, random $k$-regular networks and square lattices. 
The simulations in the upper branch started with $\lambda=0.7$ and $\rho_o=1$, while the simulations in the lower branch started with $\lambda=0.49$ and $\rho_o=1/N$. We then increase (or decrease) $\lambda$ by a constant amount $ \Delta \lambda = 0.01$ with a new initial condition corresponding to the final configuration of the preceding simulation. 
The results for 
complete graphs [Fig.~\ref{fig:hyst}(a)]
and
random $k$-regular networks [Fig.~\ref{fig:hyst}(b)] 
reveal  the presence of a hysteresis loop, \textcolor{black}{history-dependence of the  dynamics, } which is a signature of a first-order transition. These results are in agreement with the corresponding 
mean-field prediction, as expected, since 
complete graphs and random $k$-regular networks 
are infinite-dimensional systems.
On the other hand, for the simulations on the square lattice [Fig.~\ref{fig:hyst}(c)], we did not find  differences between the upper and lower branches, 
which in turn suggests the absence of hysteresis and of a discontinuous transitions for the MVE model on this two-dimensional lattice. 
\textcolor{black}{Moreover, these findings underscore the dependence of hysteresis on the dimensionality of the system.}

The sensitivity 
of the stationary order parameter 
to initial conditions 
is another test that provides clues about the nature of 
 absorbing-state phase transitions
 for a given set of parameters~\cite{sampaio2018symbiotic,assis2009discontinuous}.
The results of such a test are shown in  Fig.~\ref{fig:biestability} where we plot several time series of the infected density $\rho$ for initial conditions in the interval $[ 1/N, 1 ] $.
The simulations on full graphs
[Fig.~\ref{fig:biestability}(a)]
and 
random $k$-regular networks [Figs.~\ref{fig:biestability}(b)] 
reveal that the stationary state depends on the initial conditions in a bistable mode. Such bistability confirms that the MVE model 
undergoes a discontinuous phase transitions 
on structures with infinite dimension. 
On the other hand, on square lattices [Fig.~\ref{fig:biestability}(c)]  all the time series of $\rho$ converge to a single stationary state for all the considered initial conditions $\rho_o$. 
\textcolor{black}{These results indicate that the spatial constraints in 2D lattices possess the capacity to disrupt the bistability of a dynamic process. Such absence of bistability provides  further evidence that the phase transition of the MVE model remains  continuous in this two-dimensional structure.}

In Fig.~\ref{fig:crit-expon-2} we provide a finite-size scaling (FSS)
analysis of the MVE model on square lattices.
From now on, we set $\rho_o=1$ in all simulations.
 In Fig.~\ref{fig:crit-expon-2}(a)  we 
plot $\ln{\rho(t)}$  versus $\ln{t}$
for different values of $\lambda$ 
to monitor the deviations from the asymptotic
power-law decay $ \rho(t) \sim t^{-\delta} $ where  $\delta$ is the 
 critical density decay exponent. The positive (negative) curvature for large t indicates that the system has a tendency to reach the active (absorbing) phase. 
 From the fit with the smallest curvature we find that  
 the critical behavior in time $ \rho(t) \sim t^{-\delta} $ holds satisfactorily with 
 $\delta = 0.489\pm0.082$ and 
 $\lambda_c = 0.5898\pm 0.0005 $ for a sufficiently large value of $t$.

Still in  Fig.~\ref{fig:crit-expon-2} we employ the data collapse technique for estimating other critical exponents. 
The scaling hypothesis predicts that all 
properly scaled curves can be
collapsed onto a single curve. To perform a data collapse we invoke the following scaling laws~\cite{hinrichsen2000non},
\begin{align} 
\rho(t) & \sim t^{-\beta/\nu_{||}} \mathcal{F} ( \Delta t^{ 1/\nu_{||} }  ) \label{eq:scalingDP-1}
\\
\rho(t) & \sim L^{-\beta/\nu_{\perp}} \mathcal{G}(  t L^{ -z}  ),       \label{eq:scalingDP-2}
\end{align} 
where $\Delta =\lambda-\lambda_c $ is the deviation from the critical point, and $\mathcal{F}$ and $\mathcal{G}$ are  scaling functions.
In turn, $z$ is the so-called dynamic critical exponent, whereas $\nu_{||}$ and  $\nu_{\perp}$ are the critical exponents associated with the temporal and spatial correlation lengths, respectively.  
Based on Eq.~\ref{eq:scalingDP-1}
 we plot 
 $ \ln{ (\rho (t) t^{\beta/\nu_{||} }) } $ versus   $ \ln{ (t \Delta^{\nu_{||}}) } $ 
 for different values of $\Delta$
 as depicted in Fig.~\ref{fig:crit-expon-2}(b). 
 Scanning several candidates for  $\{\beta,\nu_{||}\}$  we find that  all curves  collapse on a single curve for 
 $\beta=0.58\pm 0.02$ and 
 $\nu_{||}=1.29\pm 0.01$.
 Based on Eq.~\ref{eq:scalingDP-2}, the scaled order parameter 
 $\rho(t) L^{\beta/\nu_{\perp}}$ is  plotted against  $ t L^{-z}$ 
 for various system sizes in the double-logarithmic graph shown in Fig.~\ref{fig:crit-expon-2}(c). By tuning   $\{z, \beta,\nu_{\perp}\}$  the data points collapse onto a single curve for a sufficiently large value of $t$, which gives the estimated values 
 $z=1.73\pm 0.12$, 
 $\beta=0.58\pm 0.05$ 
 and 
 $\nu_{\perp}=0.74\pm 0.04$.

Taking a big-picture look at the results shown in  the panels of Fig.~\ref{fig:crit-expon-2}, we note that the
independent estimates for critical exponents are statistically consistent with each other. 
The best estimated values are summarized in Table~\ref{tab:exponents_MVE_2d}. 
In many nonequilibrium models the triplet $\{\beta, \nu_{||}, \nu_{\perp} \}$ is the fundamental set of critical exponents that labels the universality class~\cite{hinrichsen2000non}. The other critical exponents can be obtained from the relations $ z = \nu_{||}/\nu_{\perp}$ and  $ \delta = \beta/ \nu_{||}$. From Table~\ref{tab:exponents_MVE_2d} it is clear that our estimated critical exponents do satisfy these   relations within the error bars. It is evident that the active-to-absorbing state transition
of the MVE model on square lattices exhibits a scaling invariance with critical exponents compatible with the continuous transition of the directed percolation universality class. 
\textcolor{black}{Moreover, the results of the 2D-MVE dynamics provides another support to the DP conjecture of Janssen~\cite{janssen1981nonequilibrium} and Grassberger~\cite{grassberger1982}, which states that 
under very general circumstances~\cite{hinrichsen2000non} (scalar order parameter, short-range interactions and no special attributes such as additional symmetries or quenched
randomness) models that exhibit a 
continuous phase transition from a fluctuating active phase into a \emph{single} absorbing state should belong to the DP universality class.}

The tricritical point (TCP) is the location in the phase diagram that separates continuous  from discontinuous transitions. At the TCP a new universality class emerges~\cite{araujo2011tricritical, cellai2011tricritical,cao2012correlated,min2018competing}. In our case, sufficiently away from the TCP we observe that the MVE model belongs to the DP universality class. Then at the TCP we should expect a universal behavior according to the Tricritical Directed Percolation (TDP) universality class~\cite{lubeck2006tricritical,grassberger2006tricritical}.
Indeed, this is what we observe in the results compiled in the panels of  Fig.~\ref{fig:tricrit-expon-0} and Table~\ref{tab:exponents_tricri}.

In Fig.~\ref{fig:tricrit-expon-0} we provide the tricritical analysis of the MVE on full graphs. 
As we did not detect discontinuous transitions in the  MVE model on square lattices, the tricritical analysis does not apply to these systems.
 Again, we start all MC simulations from $\rho_o=1$. To assist our FSS analysis, we use some exact results from the mean-field calculations, namely the location of the TCP given by Eq.~\ref{eq:tcp} with 
 $\eta=0.1$ and $\gamma=0.07$: $(\lambda_T,\sigma_T) = (0.18889, 0.62941) $.  
 In the panel (a) we show that the power-law decay $ \rho(t) \sim t^{-\delta^T} $ is statistically satisfied at $ \lambda_T $
for the estimated exponent 
$\delta^T = 0.509 \pm 0.013 $. 
Afterward, we employ a data collapse method for estimating other tricritical exponents.
In Fig.~\ref{fig:tricrit-expon-0}(b) we plot 
$\rho(t) t^{-\beta^T/\nu_{||}^T } $ versus $ t \Delta^{\nu_{||}^T} $ 
for several distances from the TCP, $\Delta=\lambda-\lambda_T$. 
We find that all curves  collapse for the set 
$\beta^T=0.50\pm0.01$ and 
$\nu_{||}^T=1.01\pm0.02$.
In Fig.~\ref{fig:tricrit-expon-0}(c) the rescaled order parameter
$\rho(t) N^{\beta^T/\nu_{\perp}^T}$ is plotted against $t/N^{z^T}$
for various system sizes. By tuning $\{z^T, \beta^T,\nu_{\perp}^T \}$ the data points collapse onto a single curve 
for a sufficiently large value of $t$, which gives the estimated values 
$z^T=1.9\pm0.2$,
$\beta^T=0.50\pm0.02$ and 
$\nu_{\perp}^T=0.49\pm0.03$.

Let us now focus on the panel (d) of Fig.~\ref{fig:tricrit-expon-0} where we perform the crossover analysis~\cite{lubeck2006tricritical}  based on 
the crossover scaling ansatz for the order parameter given by
\begin{align} 
\rho & \sim \Delta_\sigma^{\beta^T/\phi^T } \mathcal{H} ( \Delta_\lambda \Delta_\sigma^{ -1/\phi^T }  ),  \label{eq:scalingTDP}
\end{align} 
where $\Delta_\sigma = \sigma_T-\sigma $ and 
$\Delta_\lambda =\lambda-\lambda_c $. In turn,
$\mathcal{H}$ is a  scaling function and $\phi^T$ is the crossover critical exponent.
In the  Fig.~\ref{fig:tricrit-expon-0}(d), we plot the rescaled order parameter
$\rho \Delta_\sigma^{-\beta^T/\phi^T }$
versus 
$\Delta_\lambda \Delta_\sigma^{ -1/\phi^T}$. 
We see that our MC data do collapse reasonably well which, in turn,  yields the estimated values for 
$\beta^T=0.49\pm0.01$ and 
$\phi^T=0.52\pm0.02$.

Taking into account all the results shown in the four panels of  Fig.~\ref{fig:tricrit-expon-0} we note that the
independent estimates for critical exponents are statistically consistent with each other. 
The best estimated values summarized in Table~\ref{tab:exponents_tricri} 
indicate that  the relations between the tricritical exponents $ z^T = \nu_{||}^T/\nu_{\perp}^T$,  $ \delta^T = \beta^T/ \nu_{||}^T$ are satisfactorily preserved. All these results show that the tricricality of the MVE on complete graphs also belongs to the TDP universality class.

\noindent\begin{minipage}{\linewidth}
\centering

\includegraphics[width=0.455\textwidth]{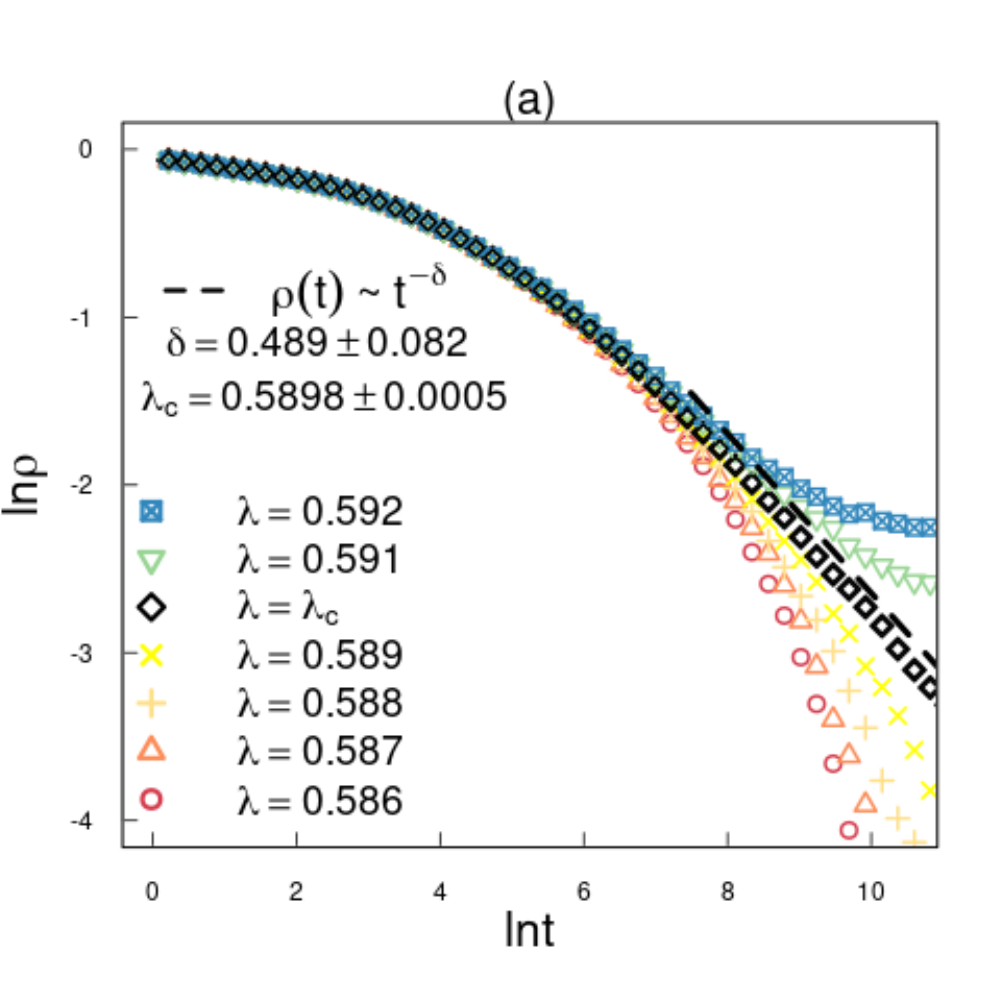}
\includegraphics[width=0.455\textwidth]{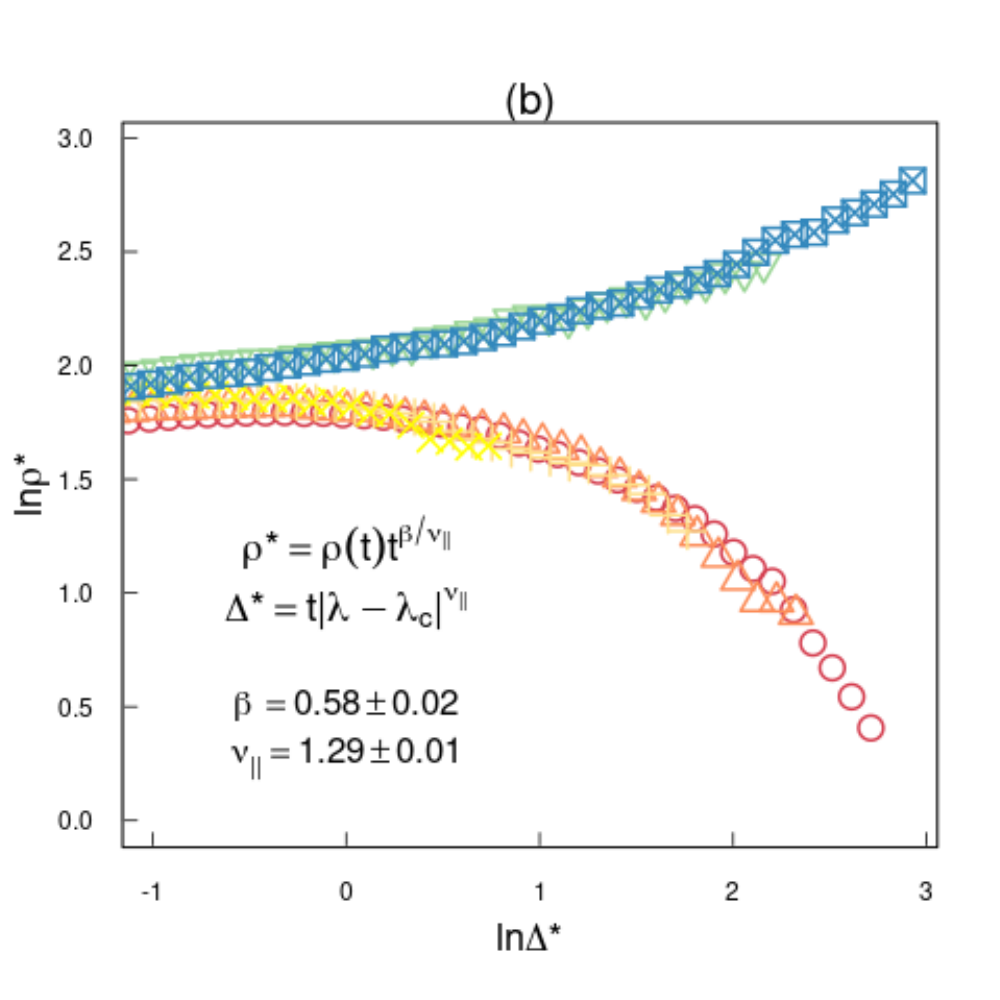}
\includegraphics[width=0.455\textwidth]{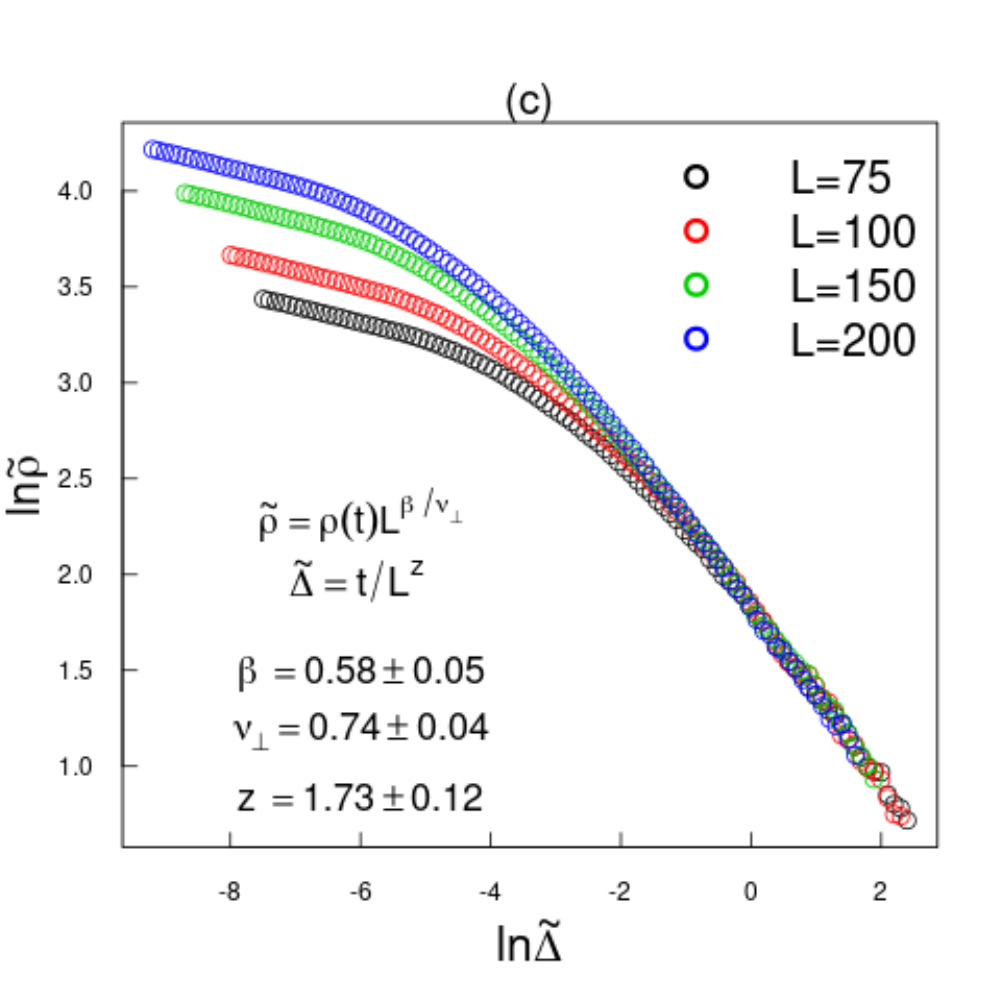}

\captionof{figure}{ Critical behavior of the MVE on square lattices. All panels are in log-log scale.
Panel (a): $\rho(t)$ versus $t$ for simulations performed with $L=100$ and several $\lambda$ around the critical point;  the black dashed line corresponds to the fit with the least curvature.
Panel (b): 
$\rho^{*} = \rho (t) t^{\beta/\nu_{||} } $ versus   $ \Delta^{*} = t (\lambda-\lambda_c)^{\nu_{||}} $ for
simulations with $L=100$; the color legend of panel (a) also applies to panel (b).
Panel (c): 
$ \Tilde{\rho} = \rho(t) L^{\beta/\nu_{\perp}}$ versus $\Tilde{\Delta} = t L^{-z}$ for simulations performed with $L=\{75,100,150,200\}$.
In panels (b-c) we use the  $\lambda_c$ estimated with the procedure shown in panel (a).
In all cases we use $\rho_o=1$ and we let the simulations run until the dynamics reaches the stationary state.
 The estimated critical quantities are shown inside each panel and the best estimates are compiled in the  Table~\ref{tab:exponents_MVE_2d}.   }
\label{fig:crit-expon-2}

\captionof{table}{ Critical exponents of the MVE obtained from the FSS analysis presented in Fig.~\ref{fig:crit-expon-2}. The exponents for the Directed Percolation (DP) universality class were taken from the appendix A.3.1 of Ref.~\cite{lubeck2004universal}. }
\renewcommand{\arraystretch}{1.5}
\begin{tabular}{c c c c c c} 
 \hline
2D Lattice \hspace{1cm} & $\beta$ & $z$ & $\delta$ & $\nu_{||}$ & $\nu_{\perp}$ \\ 
 \hline  
MVE \hspace{1cm} & $0.58\pm 0.02$    & $1.73\pm 0.12$      & $0.489\pm 0.082$   & $1.29\pm 0.01$   & $0.74\pm 0.04$ \\ 
DP  \hspace{1cm} & $0.5834\pm 0.0030$ & $1.7660\pm 0.0016$ & $0.4505\pm 0.0010$ & $1.295\pm 0.006$ & $0.7333\pm 0.0075$\\ 
 \hline
 \end{tabular}
 \label{tab:exponents_MVE_2d}
\end{minipage}

\noindent\begin{minipage}{\linewidth}
\centering

\includegraphics[width=0.495\textwidth]{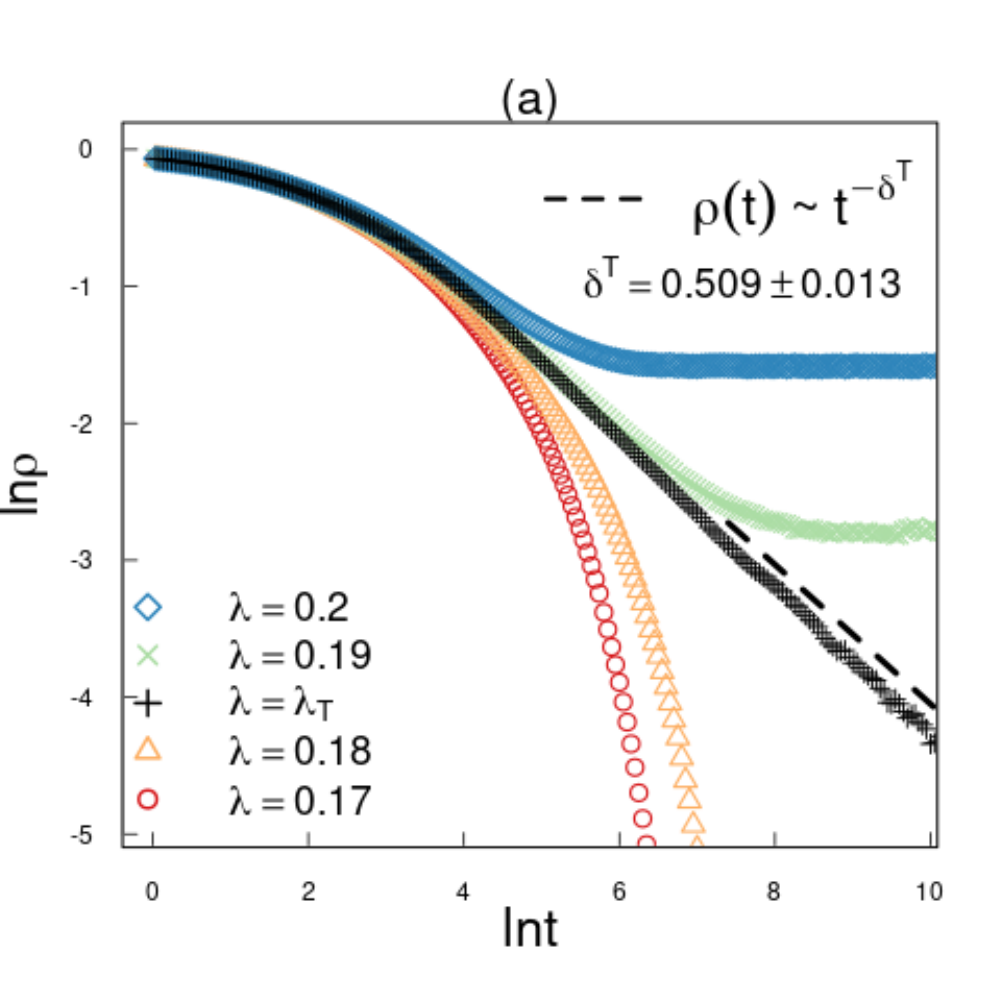}
\includegraphics[width=0.495\textwidth]{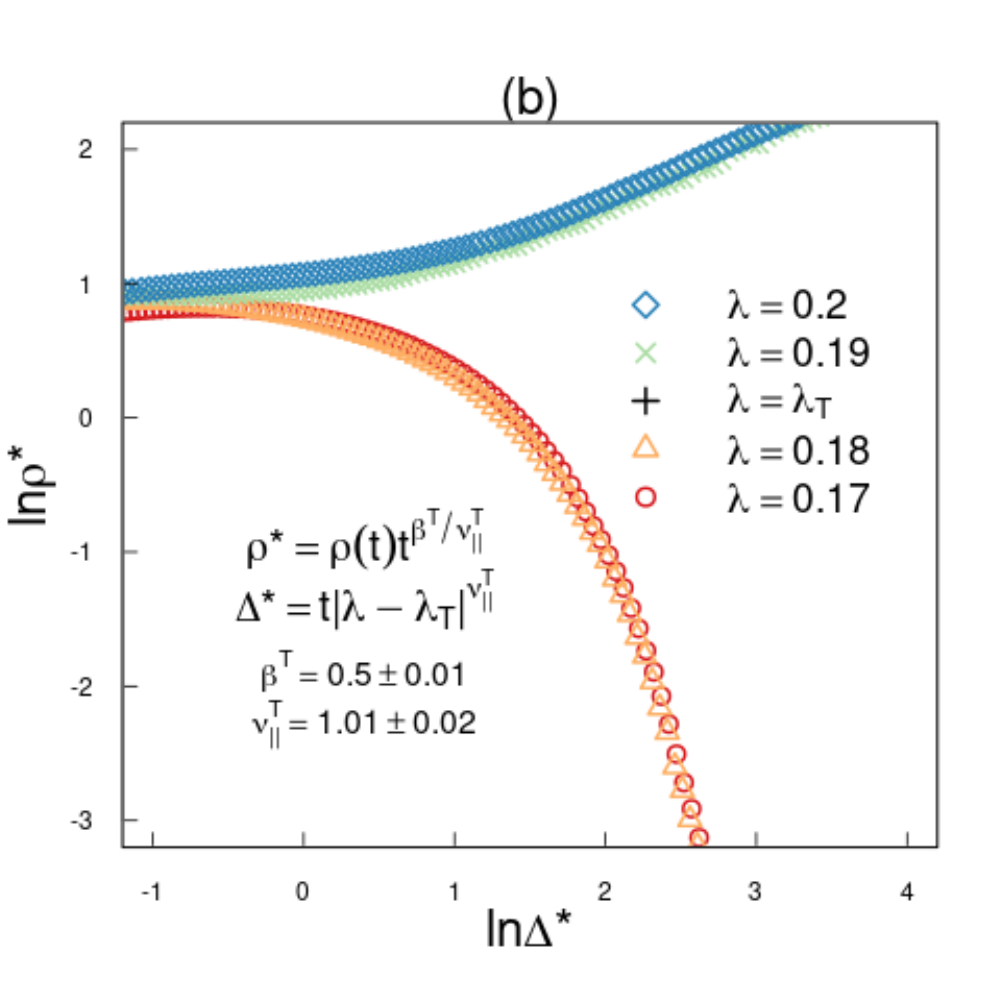}

\includegraphics[width=0.495\textwidth]{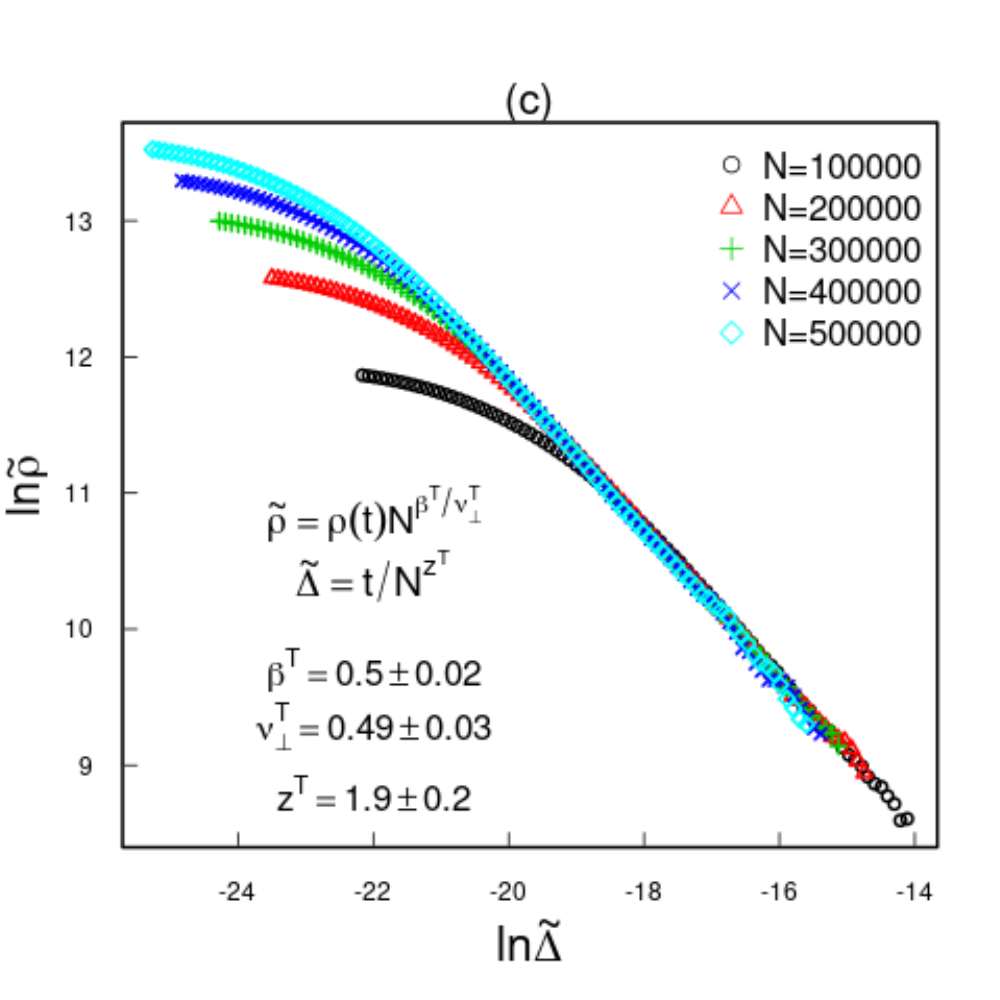}
\includegraphics[width=0.495\textwidth]{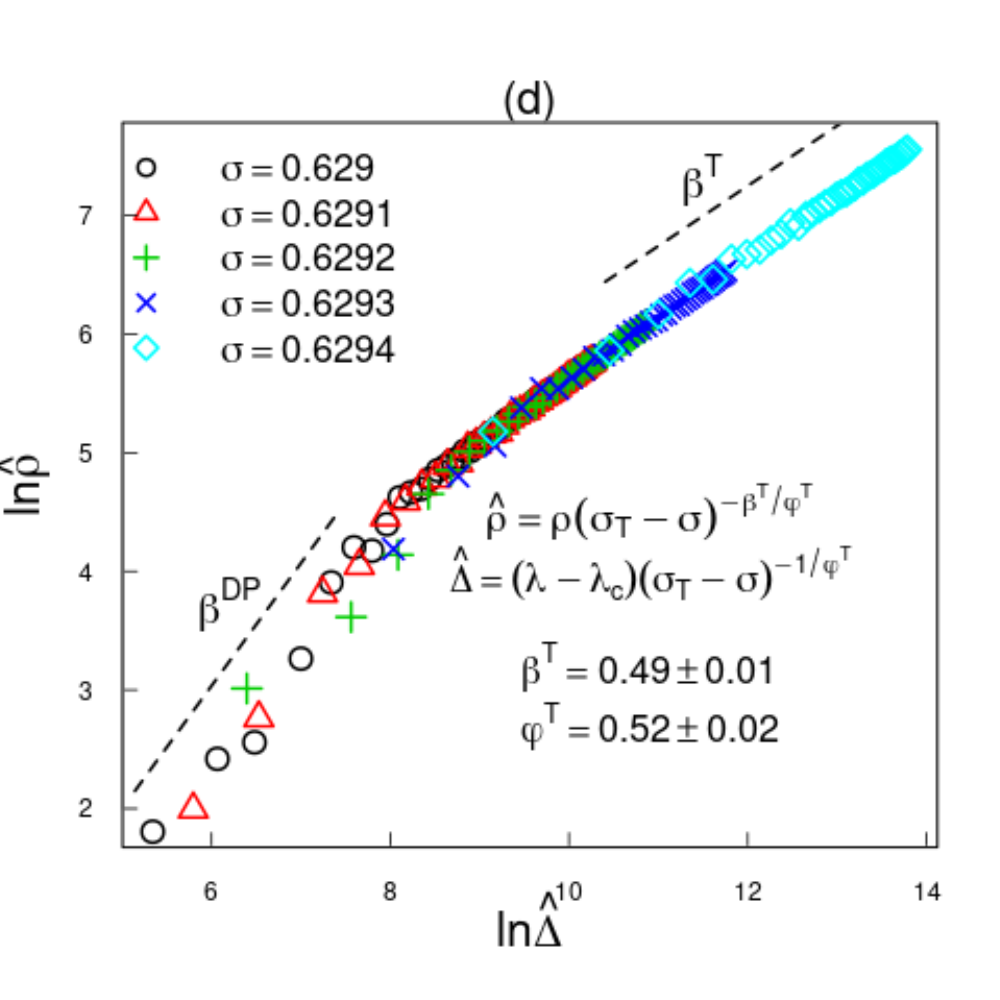}

\captionof{figure}{\small Tricritical behavior for the MVE  on complete graphs.  To assist the FSS analysis, we use the location of the TCP obtained from Eq.~\ref{eq:tcp}: $ \lambda_T = 0.18889$ and $ \sigma_T = 0.62941$. All panels are in log-log scale.
Panel (a): 
$\rho(t)$ versus $t$ for simulations performed with $N=5.10^5$ and different $\lambda$ around the TCP;  the black dashed line corresponds to the fit at $\lambda=\lambda_T$.
Panel (b): 
$\rho^{*} = \rho (t) t^{\beta^T/\nu_{||}^T } $ versus   $ \Delta^{*} = t (\lambda-\lambda_c)^{\nu_{||}^T } $ 
for $N=5.10^5$  and different $\lambda$ near the TCP; the color legend of panel (a) also applies to panel (b).
Panel (c): 
$ \Tilde{\rho} = \rho(t) L^{\beta^T/\nu_{\perp}^T}$ versus $\Tilde{\Delta} = t N^{-z^T}$
 for $N=\{10^5-5.10^5\}$ and $\lambda=\lambda_T$. 
Panel (d): 
$ \hat{\rho} =\rho (\sigma_T-\sigma)^{-\beta^T/\phi^T } $
versus $\hat{\Delta} = (\lambda-\lambda_c) (\sigma_T-\sigma)^{ -1/\phi^T}$  
 for  $N=5.10^5$; the dashed lines  are  guidelines to the eye with the exact slopes $\beta^{DP}=1$ and $\beta^{T}=1/2$.
 In all cases we use $\rho_o=1$ and we let the simulations run until the dynamics reaches the stationary state.
 The estimated values for the tricritical exponents are shown inside each panel and the best values are summarized in Table~\ref{tab:exponents_tricri}. }
\label{fig:tricrit-expon-0}

\captionof{table}{Tricritical exponents for  the MVE obtained from the FSS analysis presented in Fig.~\ref{fig:tricrit-expon-0}.  The theoretical mean-field exponents were taken from Refs.~\cite{lubeck2006tricritical,grassberger2006tricritical}.}
\renewcommand{\arraystretch}{1.5}
 \begin{tabular}{c c c c c c c} 
 \hline
 Complete Graph &  $\phi^T$  & $\beta^T$  & $\delta^T$ & $\nu_{||}^T$ & $\nu_{\perp}^T$ & $z^T$ \\ 
 \hline
 Estimated  & 0.52$\pm$0.02 & 0.50$\pm$0.01  & 0.509$\pm$0.013   & 1.01$\pm$0.02 & 0.49$\pm$0.03 & 1.9$\pm$0.2 \\ 
 Expected   & 1/2           & 1/2            & 1/2             & 1             &  1/2         &  2 \\ 
 \hline
 \end{tabular}
 \label{tab:exponents_tricri}
\end{minipage} 

\section{Final remarks}\label{sec:remaks}

 We characterized the  critical and tricritical phenomena associated with the dynamics of a  minimal vaccination-epidemic (MVE) model. Our results obtained through a mean-field approach, MC simulations and finite-size scaling (FSS) theory revealed the presence of continuous and discontinuous phase transitions depending on the dimensionality of the system.

Despite much theoretical progress, there is no general framework yet to treat the subject of first-order phase transitions in low-dimensional nonequilibrium systems. 
While this subject was clarified in 1D systems with the Hinrichsen's conjecture~\cite{hinrichsen2000first}, the full picture for 2D systems is still not clear. At present it is known that,
 discontinuous nonequilibrium  transitions into an absorbing state detected in the mean-field limit can also take place in  2D lattices~\cite{assis2009discontinuous,windus2007phase,da2011critical,da2012two,fiore2014minimal,de2015generic,de2016temporal}, but this is not always the case~\cite{sampaio2018symbiotic,chen2017fundamental}.
For instance, recently it was shown~\cite{sampaio2018symbiotic} that the phase transition of the two-dimensional symbiotic contact process~\cite{de2012symbiotic}   is  continuous,  in disagreement with the discontinuous transition observed in the mean-field limit.
Here, we show that the MVE model undergoes a first-order absorbing state transition on systems with infinite dimension, namely complete graphs and random $k$-regular networks. However, for the MVE model on regular square lattices, we find no signs of a discontinuous transition to the absorbing state.

Previous studies have introduced stochastic extensions  of the SIS  model capable of generating discontinuous transitions~\cite{chen2017fundamental,bottcher2017critical,dodds2004universal,gross2006epidemic,bottcher2015disease,chae2015discontinuous,gomez2016explosive,chen2017epidemic,2018piresOC,matamalas2020abrupt}. But none of them  presented the possibility of universal features emerging  at the line separating the regime of discontinuous and continuous transitions. Here, we also show  that the 
tricritical point of the MVE on full graphs is  associated with scaling laws compatible with  an independent universality class, namely the Tricritical Directed Percolation (TDP)~\cite{lubeck2006tricritical,grassberger2006tricritical}.
It is important to note that  our tricritical analysis is different from the one presented in Ref.~\cite{janssen2004generalized}, where the authors studied
the so-called generalized general epidemic process (GGEP)  which is an extension of the SIR model with four classes of
individuals. They reported that the GGEP can be tuned to fall in the universality class of the Tricritical Dynamic Isotropic Percolation (TDIP).

\textcolor{black}{ In summary, our work adds statistical and spatial insight into epidemic-vaccination models that present nonequilibrium continuous and discontinuous phase transitions. }  
In future works,  it would be interesting to investigate the critical phenomena arising from the MVE model in a low-dimensional system of mobile agents. 
As documented in previous studies~\cite{gonzalez2004scaling,villa2015eluding,pianegonda2015effect,de2017effects,polovnikov2022subdiffusive}, diffusion can play multiple and contrasting roles in nonequilibrium models that  have  active-to-absorbing phase transitions.

\section*{Acknowledgments}
We gratefully acknowledge CNPq, CAPES, FUNCAP and the National Institute of Science and Technology for Complex Systems in Brazil for financial support.

\bibliography{main.bib}

\end{document}